\documentclass[a4paper,12pt]{iopart}
\usepackage[utf8]{inputenc}
\usepackage{cite,epsfig}
\usepackage{graphicx}
\usepackage{amssymb}
\usepackage{iopams}
\usepackage{color}

\newcommand{\ud}{\mathrm{d}}

\newcommand{\shh}[0]{SHIELD-HIT}
\newcommand{\swa}[0]{$S_{\mathrm{water/air}}$}
\newcommand{\rr}[0]{$R_{\mathrm{res}}$}
\newcommand{\rp}[0]{$R_{\mathrm{p}}$}
\newcommand{\ea}{\emph{et al.}}

\begin{document}

%
%
%
\title[Analytical expressions for water-to-air stopping-power ratios]
{Analytical expressions for water-to-air stopping-power ratios relevant for 
accurate dosimetry in particle therapy}

\author{Armin L\"uhr$^{1,2}$, David C.\ Hansen$^{2}$, Oliver J\"akel$^{3,4}$,
 Nikolai Sobolevsky$^{5}$, 
 and Niels Bassler$^{1,2}$}


\address{$^1$ Department of Experimental Clinical Oncology, Aarhus
  University Hospital, Aarhus, Denmark}
\address{$^2$ Department of Physics and Astronomy, University of Aarhus, Aarhus,
Denmark}
\address{$^3$ Department of Medical Physics in Radiation Oncology, German
  Cancer Research Center (DKFZ), Heidelberg, Germany}
\address{$^4$ Heidelberg Ion Beam Therapy Center (HIT), Heidelberg
  University Hospital, Heidelberg, Germany}
\address{$^5$ Department of Neutron Research, Institute for Nuclear Research of the Russian Academy of Sciences, Moscow 117312, Russia}

\ead{luehr@phys.au.dk; bassler@phys.au.dk}

\section*{Abstract}
\begin{abstract}
 In particle therapy, knowledge of the stopping-power ratio (STPR) 
 of the ion beam for water and air is necessary for accurate
 ionization chamber dosimetry. Earlier work has investigated the STPR
 for pristine carbon ion beams, but here we expand the calculations
 to a range of ions ($1\le z \le 18$) as well as spread out Bragg peaks
 (SOBPs) and provide a theoretical in-depth study with a special focus on the 
 parameter regime relevant for particle therapy.

 The Monte Carlo transport code \shh\ is used to calculate complete
 particle-fluence spectra which are required for determining STPR
 according to the recommendations of the International Atomic Energy Agency
 (IAEA).
%

%
 The STPR at a depth $d$ depends primarily on the average energy of the primary
 ions at $d$ rather than on their charge $z$ or absolute position in the medium.
 However, STPRs for different sets of stopping-power data for water and air
 recommended by the International Commission on Radiation Units \&
 Measurements (ICRU) are compared, including also the recently revised data
 for water, yielding deviations up to 2\% in the plateau region. 
 In comparison, the influence of the secondary particle spectra on 
 the STPR is about two orders of magnitude smaller in the whole region 
 up till the practical range. 
 The gained insights enable us to propose simple analytical expressions for the 
 STPR for both pristine and SOBPs 
 as a function of penetration depth depending parametrically on the practical
 range. 
%

\end{abstract}

%
\maketitle


%
%

\section{Introduction}
\label{sec:intro}
Stopping powers are essential for calculating the dose deposited by
ionizing particles. The deposited dose is described as the mass
stopping power multiplied with the particle fluence, 
while assuming charged particle equilibrium from the short-ranged delta
electrons. At particle-therapy centers air-filled ionization chambers are
routinely used as a main tool for quality assurance of the delivered
beam. 
Several dosimetry protocols for protons have been conceived while 
the most recent protocol provided by the 
International Atomic Energy Agency (IAEA), 
TRS-398~\cite{trs398}, 
sets the standard in proton dosimetry today. 
In addition, TRS-398 also covers dosimetry for ions heavier than protons. 
The protocol uses an absorbed dose-to-water based formalism
and relates the dose to water $D_{w,Q}$ to the acquired charge $M_Q$
multiplied by a calibration factor $N_{D,w,Q_0}$ and a dimensionless beam quality
correction factor $k_{Q,Q_0}$. The correction factor $k_{Q,Q_0}$ 
relates the measured beam quality $Q$ to the beam quality $Q_0$ used for 
calibration of the dosimeter 
%
and it is defined in TRS-398 as
\begin{equation}
  \label{eq:k_factor_definition}
  k_{Q,Q_0} = \frac{(S_{\mathrm{water/air}})_Q}{(S_{\mathrm{water/air}})_{Q_0}}
            \frac{(W_{\mathrm{air}})_Q}{(W_{\mathrm{air}})_{Q_0}}
            \frac{p_Q}{p_{Q_0}}\,
\end{equation}
including the water-to-air stopping-power ratio, \swa , the 
mean energy expended in air per ion pair formed, $W_{\mathrm{air}}$,
and a perturbation factor $p_Q/p_{Q_0}$ which considers effects for 
the specific ionization chamber used.

As mentioned in TRS-398, calculating the correct beam quality factor
in particle therapy is complex since it involves knowledge of the
entire particle-energy spectrum at the point of interest. 
Instead, TRS-398 proposes a pragmatic 
approach by recommending a fixed value of 1.13 as a generic correction 
factor for the dosimetry of ions heavier than protons based on the analysis 
by Hartmann \ea~\cite{hartmann99}, irrespective of the 
particle types and energy spectra
which are functions of depth. 
Accordingly, TRS-398 summarizes that the estimated combined standard 
uncertainty in $k_{Q,Q_0}$ in ion beams heavier than protons (about 3\%) 
arises largely from the uncertainty of the stopping-power ratio (STPR) 
(about 2\%) and the value for $W_{\mathrm{air}}$ (about 1.5\%).%
\footnote{Although the study of the value for $W_{\mathrm{air}}$ is beyond the scope of this work it shall be mentioned that the estimated uncertainty for $W_{\mathrm{air}}$ calls for a detailed investigation.} 
This has been taken up by Henkner \ea~\cite{henkner09} and Geithner
\ea~\cite{paul07,geithner06} for mono energetic \emph{carbon} beams,
and they found out that (i) the STPR is not constant but varies with
pentration depth, and (ii) it depends strongly on the accuracy of the
stopping-power data used as input for the calculation. Accordingly, it
was concluded that in a clinical setting an over or under dosage may
occur in the order of a few percent.

Here, we shall continue the initiated work on STPRs focusing on two objectives.
First, gaining a sound understanding of the physics determining the STPR, 
and second, exploiting the gained understanding in order to provide results 
with direct clinical relevance which are ready to be applied in clinical 
practice in a quality assurance setting.

The deeper insight in the context of STPR is required since STPR strongly 
depends on the stopping-power data which are used for its determination. 
The problem is, however, that the stopping-power data currently recommended 
by the International Commission on Radiation Units and Measurements (ICRU) 
possess some intrinsic inconsistencies. In contrast to an accurate but purely 
numerical calculation of STPRs, a sound understanding of the relevant physics 
allows for conlclusions independent of the employed set of stopping-power data.
It is also a prerequisit for an analytical description of the STPR. 
It should be emphasized that, in contrast to earlier work on STPR 
\cite{henkner09,geithner06,paul07}, the present study also considers 
the recently revised ICRU 73 \cite{icru73} stopping-power data for ions 
heavier than helium on water.
These data replace the ones originally published in ICRU 73 which led to a 
still ongoing discussion on stopping powers for water targets 
(see, e.g., \cite{paul10}) especially in view of recent measurements 
by Schardt \ea\ \cite{schardt08}.


%


TRS-398 explicitly states that the STPR 
for water-to-air, \swa , should be obtained by averaging over the complete 
spectra of particles present. And consequently, this requirement was 
considered to be an important limitation in the case of heavy charged 
particles, where the determination of all possible particle spectra was 
assumed to be a considerable undertaking.
This was certainly the case a decade ago and may still be true 
from the point of view of dose determination for routinely quality assurance.
Nowadays, however, the determination of complete spectra of particles 
can be achieved conveniently and with high accuracy by applying Monte Carlo 
transportation codes exploiting the commonly available computer power. 
These codes are in general valuable in predicting radiation fields of 
ions in tissues and are in particular useful in hadron therapy for the 
simulation of ion transport. 
The most common codes in particle therapy with ions heavier than protons are  
Geant4 \cite{agostinelli03},  FLUKA \cite{fasso05}, PHITS \cite{iwase02}, 
MCNPX \cite{pelowitz05}, and  \shh \cite{dementyev99,gudowska04},
all taking into account the atomic interaction of the ions with the
target medium  
as well as the nuclear interaction. 
It is the former interaction which mainly determines the energy loss of 
the incident ions and therefore the stopping power, while the latter 
interaction is responsible for fragmentation and therefore for the production
of secondary-particle spectra.

Initial studies on STPRs relevant for dosimetry in radiation therapy with 
ions heavier than protons were 
performed without Monte Carlo calculations ignoring the influence of the
secondary particle spectrum (e.g.\ \cite{salamon80,hiraoka95,icru49} as 
presented in TRS-398 \cite{trs398}).
Calculations exploiting the capabilities of Monte Carlo codes were performed 
with \shh\ but exclusively carbon-ion fields 
\cite{geithner06,paul07,henkner09} were studied. 
However, the dependence of the STPR on different ion species is of
intrest since a number of facilities world-wide (e.g., NIRS and HIT)
are equipped with radiation fields which cover a broader range of ions
than merely protons and carbon ions. Furthermore, it was recently argued 
that ions heavier than carbon may play an important role in the near future 
concerning the radiation therapy of radio-resistent tumors \cite{bassler10b}. 
Consequently, a large variety of ion species, namely, H, He, Li, C, N, O, Ne, 
Si, and Ar are considered here --- all accessable either for clinical 
radiation therapy (up till O and Ne at HIT and NIRS, respectively) or for 
in vitro radiobiology experiments.

Despite their obvious relevance in medical application, so far, STPRs for 
spread-out Bragg peaks (SOBPs) for ions heavier than protons have been 
discussed only scarcely in literature, namely, by Henkner \ea\ 
\cite{henkner09}. 
In the case of proton beams more detailed efforts have been performed, 
e.g., by Palmans \ea\  \cite{palmans98,palmans02} and earlier already 
by Medin and Andreo \cite{medin92}.  
Henkner \ea , who considered carbon ions, outlined in the conclusions of 
\cite{henkner09}
that a more detailed analysis of STPRs
for SOBPs is clearly needed since their statements were only based on
the analysis of a single \emph{physically} optimized SOBP using one set of
stopping-power data. 
Consequently, one focus of this work should be a systematic study of the
STPR for SOBPs, both \emph{physically} and \emph{biologically} optimized,
of different widths and practical ranges 
leading to an analytic expression for the  STPR.

%

This paper is organized as follows:
%
First, the physics relevant for the STPR and 
the employed methods are discussed. Furthermore, analytical expressions
for the average energy of the primary ions and STPRs are proposed.
Subsequently, the results for the water-to-air STPR for 
pristine as well as SOBPs are presented
and compared to the proposed analytical expressions.
The following discussion concentrates on three issues, namely, the influence 
of the stopping-power data on the STPR, the dependence of the STPR on the ion 
energy, and STPRs for SOBP. 
%


%

\section{Materials and Methods}
\label{sec:mm}

\noindent For all our calculations we used the Monte Carlo particle transport
code \shh\ \cite{gudowska04,geithner06}, based upon the most recent version
\shh08 \cite{shield_url}. 
A number of improvements and new functionalities were added to \shh08, 
doc\-u\-mented in \cite{hansen11a}, finally resulting in \shh10A 
\cite{hansen11}. 
Here, only the relevant changes are reported. 
First, there is now the possibility of directly scoring the 
STPR of any media, described in detail in 
section~\ref{sec:mc_scoring}.
Apart from this, raster scan files generated by the treatment planning
software TRiP~\cite{kraemer00,kraemer00a} can now be read by \shh\
in order to recalculate SOBPs. 
In this study, we present calculations from four single field carbon ion SOBPs,
listed in Table~\ref{tab:SOBP}. The width of the SOBP is defined as usual 
by the width in which the dose is above 95\% percent \cite{trs398}. 
All SOBPs are 3-dimensional dose cubes with equal side lengths. 
The resulting raster-scan file describes
the needed amount of particles for each raster point and for each energy
slice providing the necessary input for \shh\ to
generate the radiation field for the SOBP.
%
A ripple filter implementation based on the design described by Weber
\ea~\cite{weber99} is added to \shh\ in a similar way as specified
by Bassler \ea~in\cite{bassler10}, in order to produce flat SOBPs.

The practical range, \rp , is  defined for protons as the depth at which
the absorbed dose beyond the Bragg peak or SOBP falls to 10\% of its
maximum value \cite{trs398}. However, for ions heavier than protons this 
definition of \rp\ 
is not feasible due to the pronounced dose tail of secondary  
particles. 
Therefore, the depth at which the absorbed dose beyond the Bragg peak or
SOBP decreases to 50\% of its maximum value is proposed and used here for 
ions heavier than hydrogen, i.e., $z>1$. 
Also other definitions of \rp\ have been used before as discussed 
in \cite{kempe08}. 
The residual range \rr\ at a depth $d$ is than defined as 
\begin{equation}
 \label{eq:residual_range_definition_TRS}
  R_{\mathrm{res}} = R_{\mathrm{p}} - d
\end{equation}
%
%
and the measurement depth $d_{\mathrm{ref}}$ 
at the middle of the SOBP in accord with TRS-398 \cite{trs398}.

\begin{table}[b]
  \centering
  \caption{Specifications of four spread out Bragg peaks (SOBPs) for carbon
    ions in water. 
    Given are the width along the beam  axis, the practical range \rp , 
    and whether the SOBP is optimized for a homogeneous physical dose or 
    relative biological dose.  The optimization 
    was performed by the treatment planning program 
    TRiP \cite{kraemer00,kraemer00a}. 
    \label{tab:SOBP}} 
  \begin{tabular}[bt]{crccc}
    \hline
    \hline
    SOBP & Width (mm) & $R_p$ (mm) & Optimization \\
    \hline 
    $a$ &  50 \quad \ & 220 & physical   \\
    $b$ &  80 \quad \ & 168 & physical   \\
    $c$ &  50 \quad \ & 150 & physical   \\
    $d$ & 100 \quad \ & 153 & biological \\
    \hline
    \hline
  \end{tabular}
\end{table}
%


\subsection{Stopping powers and mean excitation energy}
\label{sec:stopping_power}

Stopping power $S$ is defined as the average energy change $\ud E$ of a 
particle per unit length $\ud l$ in a medium. 
%
At high energies, that is about from 10 MeV/u up to 1 GeV/u,%
\footnote{The energy regime from  10 MeV/u up to 1 GeV/u, corresponds 
          according to the revised tables for water in ICRU 73 
          \cite{icru73}, for carbon ions 
          to a range from 0.0427 cm up to 108.6 cm.} 
the mean energy
loss of a charged particle to atomic electrons is well approximated by
Bethe's original theory \cite{bethe30,bethe32} which treats the
electromagnetic interaction in first-order quantum perturbation theory. 
At lower energies, however, additional higher-order terms are required
in order to reproduce experimental results. The transition from the regime 
of quantum perturbation theory to the one permitting a classical treatment 
is described in Bohr's distinguished survey paper \cite{bohr48}.

A widespread formulation of Bethe's theory summarizing all terms of the 
lowest-order stopping number $L_0$ was proposed by Fano \cite{fano63}  
\begin{equation}
 \label{eq:bethe_formula_Fano}
  \frac{S}{\rho} = \frac{4\pi e^4}{m_e v^2}\frac{1}{u}\frac{Z}{A}z^2
                   \left[
		    \ln\frac{2m_e v^2}{I} +\ln\frac{1}{1-\beta^2} -
		    \beta^2 - \frac{C}{Z} - \frac{\delta}{2}
		   \right]\,,
\end{equation}
%
%
%
%
In Eq.\ (\ref{eq:bethe_formula_Fano}),  $\rho$ is the density of the medium, 
$m_e$ the electron mass, $e$ and $u$ are the elemental units of 
electric charge and atomic mass, respectively, $Z$ and $A$ are the
atomic number and the relative atomic mass of the target medium,
respectively, $v$ and $z$ are the velocity and the charge of the
projectile, and $\beta=v/c$ where $c$ is the velocity of light in vacuum.
The mean excitation energy of the target medium is denoted by $I$, while
$C/Z$ and $\delta/2$ are the shell corrections and the density-effect 
correction%
,
respectively. The second and third term in the square brackets containing 
$\beta$ originate from Bethe's relativistic extension \cite{bethe32}
and are often referred to as relativistic corrections.
The expression in Eq.\ (\ref{eq:bethe_formula_Fano}) is consistent
with the first term $L_0$ of the stopping number $L$ 
in ICRU report 49 \cite{icru49}.
%
%
For low energies the description of the stopping powers becomes more 
complicated and higher-order terms of the stopping number $L$ have to be 
taken into account in order to correct for a number of different effects, 
such as the Barkas and the Bloch correction, $L_1$ and $L_2$, respectively.  
An effective description 
for the energy regime below 
the stopping-power maximum was provided by Lindhard and Scharff 
\cite{lindhard61}
assuming 
a rise of the stopping power which is proportional to
the square root of the particle energy.

\label{sec:mean_excitation_energy}

The mean excitation energy, $I$, is a property of the medium which 
enters lo\-ga\-rith\-mi\-cal\-ly in the stopping formula Eq.\
(\ref{eq:bethe_formula_Fano}), and is responsible for most of the target 
material dependence of the stopping-power. 
It is, on the other hand, completely independent of the properties of the
projectile.
According to Eq.\
(\ref{eq:bethe_formula_Fano}), a larger $I$-value results in a smaller stopping
power and consequently in a larger range of an ion in the medium.
%
%
%
%
%
The $I$-values in the ICRU report 49 \cite{icru49} for protons and
alpha particles (retained from ICRU report 37 for electrons
\cite{icru37}) were mainly taken from measurements. In ICRU
report 73 \cite{icru73}, however, the $I$-values are mostly determined
theoretically. 
As a result different $I$-values for the same material are recommended in 
ICRU reports 49 and 73. 
Obviously, this is
inconsistent, since an $I$-value should 
not depend on the projectile. The differences existing between the ICRU reports
highlight that the accuracy of the current employed methods to determine 
stopping-power data have still to be 
improved in order to provide a consistent target description. 
\begin{table}
  \centering
  \caption{Specifications for 6 sets of stopping-power data used in this work. 
    The stopping-power data for the first three sets are determined
    internally by \shh\ (cf.\ Sec.\ \ref{sec:mc_stp}) using the given values 
    for $I_{\mathrm{water}}$ and $I_{\mathrm{air}}$. 
    while those for the sets 4 to 6 are directly read by \shh\ as text files in 
    tabulated form. For the latter, two different tables per set are used 
    distinguishing between the lightest (H and He) and heavier ions.
    The table specifies for each set its number,  $I_{\mathrm{water}}$ and 
    $I_{\mathrm{air}}$ in eV, the range of ions for which these data 
    are applied, references, and if adequate additional comments. 
    Further explanations can be found in the text.
    \label{tab:STP_sets}} 
  \begin{tabular}[b]{cllcll}
\\[-0.25cm]
    \hline
    \hline
    \multicolumn{6}{c}{\shh\ calculates stopping-power data using $I$-values}\\[0.1cm]
    Set $\#$ & $I_{\mathrm{water}}$ & $I_{\mathrm{air}}$ & ion range &
    Reference & Comments \\ 
    \hline\\[-0.25cm]
    1& 78   & 82.8 & $z\ge 1$   &  ICRU 73 \cite{icru73,icru73a} & 
    using revised $I_{\mathrm{water}}$ \cite{icru73a}\\ 
    2& 75   & 85.7 & $z\ge 1$   & ICRU 49 \cite{icru49}  & \\
    3& 80.8 & 85.7 & $z\ge 1$   & Henkner \ea\ 
    \cite{henkner09}&  \\[0.25cm]
    \hline
    \hline
    \multicolumn{6}{c}{\shh\ directly uses tabulated stopping-power data}\\[0.1cm]
    Set $\#$ & $I_{\mathrm{water}}$ & $I_{\mathrm{air}}$ & ion range &
    Reference & Comments \\ 
    \hline\\[-0.25cm]
    4& 78   & 82.8 & $z >  2$  & ICRU 73 \cite{icru73,icru73a} & 
    revised data for water \cite{icru73a}\\ 
     & 75   & 85.7 & $z\le 2$  & ICRU 49 \cite{icru49} & \\
    5& 67.2 & 82.8 & $z >  2$  & ICRU 73 \cite{icru73} & 
    only original data for water\\  
     & 75   & 85.7 & $z\le 2$  & ICRU 49 \cite{icru49} & \\
    6& 75   & 85.7 & $z =  1$  & ICRU 49 \cite{icru49} & \\
     & 75   & 85.7 & $z >  1$  & MSTAR   \cite{paul03} & 
    charge scaling of ICRU 49\\
    \hline
    \hline
  \end{tabular}
\end{table}

\subsection{Stopping powers in \shh }
\label{sec:mc_stp}

In the current implementation of \shh\ the compilation of required
stopping-power data can be done in two ways which can be chosen
independently for each target medium. 
First, stopping-power data can be calculated internally 
by \shh\ using a modified Bethe formula at high energies and a 
Lindhard-Scharff description \cite{lindhard61} at low energies for any
kind of material composition using the corresponding material-specific
values for $I$, $Z$, and $A$ as discussed before in Sec.\ 
\ref{sec:stopping_power}.
Second, an arbitrary 
stopping-power table may be read in as a formatted text file
allowing for the use of, in principle, any stopping-power data which can
be provided in electronic form. In this work the common open source library 
\emph{libdEdx} \cite{libdedx,luehr11a} which is available online is applied 
in order to provide tabulated data in formatted form from the ICRU reports 49 
\cite{berger05,icru49} and 73 \cite{icru73,icru73a} as well as MSTAR 
\cite{paul03}.


%
The Bethe formula used by \shh\ is similar to the formulation in Eq.\
(\ref{eq:bethe_formula_Fano}). But, so far no shell corrections $C/Z$ have been
considered. These are known to be most relevant for low
energies where, however, the Lindhard-Scharff description is used
instead in \shh . Furthermore, it was demonstrated that for low energies 
(about 1 MeV/u) 
the accuracy of stopping-power data is insignificant for particle 
therapy \cite{elsaesser09}.
The same argument holds for the higher-order term $L_1$.
%
Additionally, the Bethe formula is modified in order to allow
for electron capture (significant for low energies)
by using an effective energy-dependent scaling of the
projectile charge $z$ by Hubert \ea~\cite{hubert89}.
Currently, relativistic corrections proposed by Lindhard and 
S{\o}rensen \cite{lindhard96} are still missing in \shh . 
Their importance increases 
for heavy ions with large nuclei which cannot be approximated 
as point-like particles. Although their relevance for particle 
therapy should  be studied no significant impact has been expected 
so far.

Due to existing inconsistencies in the stopping-power data recommended by ICRU 
--- discussed in Sec.\ \ref{sec:mean_excitation_energy} --- 
different sets of stopping powers are used in this work, all listed in 
Table \ref{tab:STP_sets}. 
Thereby, sets 1 and 2 as well as sets 4 and 5 are 
directly related to ICRU reports. For comparison, in set 3 the preferred 
$I$-values of Henkner \ea\ \cite{henkner09} are used while set 6 employes 
the frequently used data provided by MSTAR \cite{paul03}.
The intended purpose of the sets 1 and 2 is the attempt to describe the 
target media consistently with only one $I$-value for all ions, both 
with $z\le 2$ as well as $z>2$, applying  \shh 's internal 
routine to determine the stopping power. Accordingly, set 1 uses 
only the $I$-values from ICRU report 73, $I_{73}$, (the revised value for water, 
$I_{\mathrm{water}}=78$ eV, was very recently published in the erratum to ICRU 73 
\cite{icru73a}) while only $I$-values from ICRU 49, $I_{49}$ are used in set 2. 
%
%
%
The motivation for sets 4 and 5, on the other hand, is the direct application 
of the recommended tabulated data which can be found in
ICRU reports 49 and 73 for ions with $z\le 2$ and $z > 2$, respectively.
While set 4 uses the recently revised stopping-power data for heavy-ions 
on water, set 5 uses, for comparison to earlier studies of the STPR, the water 
data as originally published in ICRU 73. 
Note, the recently revised data from ICRU 73 \cite{icru73a} were not employed 
by Henkner \ea~\cite{henkner09}.
%

\subsection{Stopping-power ratio}
\label{sec:stopping_power_ratio}

The stopping-power ratio $S_{a/b}$ between medium $a$ and medium $b$ 
is (cf.\ TRS-398%
\footnote{In IAEA TRS-398 only the water-to-air STPR is explicitly defined, 
i.e., $a$=water and $b$=air. However, this definition is also useful for 
other media combinations.}
\cite{trs398}) given as a particle fluence weighted 
average over all primary and secondary particles. It is determined by
calculating the dose ratio via track-length fluence $\Phi_{a,i}(E)$ of
particle $i$ in medium $a$ as function of particle energy $E$ and
mass stopping power $S_i(E)/\rho$ 
\begin{equation}
 \label{eq:stopping_ratio_fluence}
  S_{a/b} = \frac{\sum_i \int_{E_{\mathrm{min}}}^{\infty}
                           \Phi_{a,i}(E)\, (S_i(E)/\rho)_a\, \ud E}
			   {\sum_i \int_{E_{\mathrm{min}}}^{\infty}
                           \Phi_{a,i}(E)\, (S_i(E)/\rho)_b\, \ud E}\,.
\end{equation}
In Eq.\ (\ref{eq:stopping_ratio_fluence}) numerator and
denominator are equal except for that the mass stopping power of medium $a$
enters in the numerator and of medium $b$ in the denominator. 
%
An energy cutoff $E_{\mathrm{min}}>0$ may originate, e.g., from 
the chamber geometry. 
The contribution of ``track-ends'' to the total dose deposition and to 
the corresponding STPR was studied in  \cite{geithner06a}. There it was 
concluded 
that they are not of relevance for light-ion dosimetry which is 
in contrast to electrons, where the contribution to the total deposited 
dose can be between  6\% and 8\% \cite{icru37}. 

In contrast to the correct definition for the STPR of an ion field in Eq.\ 
(\ref{eq:stopping_ratio_fluence}), 
the \emph{ratio of stopping powers} for media $a$ and $b$ for one particle 
species of energy $E$,
\begin{equation}
 \label{eq:stopping_ratio_approximation}
  \frac{(S(E)/\rho)_a}{(S(E)/\rho)_b} = 
  \frac{\left\langle \frac{Z}{A} \right\rangle _a}
       {\left\langle \frac{Z}{A} \right\rangle _b} 
  \frac{\ln[2 m_e v^2 / I_a]}{\ln[2 m_e v^2 / I_b] }\,,
\end{equation}
has often been considered as an approximation to the STPR, e.g.\ 
in \cite{henkner09,paul07,trs398}.
The right hand side of Eq.\ (\ref{eq:stopping_ratio_approximation}) 
is expressed by Bethe's stopping formula as given 
in Eq.\ (\ref{eq:bethe_formula_Fano}) but omitting corrections.
%
%
Note, the ratio in Eq.\ (\ref{eq:stopping_ratio_approximation}), which 
considers only one particle species, is a 
function of the particle \emph{energy} $E$ in contrast to the STPR in Eq.\ 
(\ref{eq:stopping_ratio_fluence}) which has a \emph{spatial} dependence and
takes the full energy spectra of all particles into account.

\subsection{Scoring of STPR in \shh }
\label{sec:mc_scoring}

STPRs have already been obtained with \shh\ before \cite{henkner09,geithner06} 
and only the conceptual improvements in this work are discussed in the 
following.
%
The concept of virtual scoring has been introduced which now allows 
for a parallel detector geometry independent of any physical geometry. 
Therefore, there is no longer a need for introducing artificial physical 
geometries which lead to additional region boundaries. 
%
Furthermore, the STPRs are now determined \emph{on-line}, that is, 
during the transport of the particles. 
An on-line calculation has the advantage that possible influence on the 
result due to the number and size of the energy scoring and energy spacing 
is avoided.  Additionally, higher accuracy in scoring of tracks-ends can be 
achieved in principle.

The detector for the STPR resembles Eq.\ (\ref{eq:stopping_ratio_fluence}) and 
is implemented in the following way. 
When a particle traverses a bin of the STPR detector its track-length 
fluence within the bin is 
scored and directly multiplied with  $(S/\rho)_a$
of the medium 
$a$ in which the particle moves for the energy 
$(E_{\mathrm{in}} + E_{\mathrm{out}}) / 2$. $E_{\mathrm{in}}$ and $E_{\mathrm{out}}$ 
are the energies of the particle when it enters and leaves the bin, 
respectively.
Additionally, the 
\emph{same} track-length fluence is multiplied with $(S/\rho)_b$ 
of the same particle.
Both quantities are summed up individually including all 
particles passing the bin. After a full Monte Carlo 
transport simulation the two sums are divided yielding the STPR for this bin.

In this work a transport cutoff of 0.025 MeV/u is used by \shh\ which 
means that all particle tracks end once the particle energy becomes smaller.%
\footnote{This is consistent with ICRU 73 \cite{icru73} where the range tables
for liquid water show the average path length travelled for slowing-down from 
initial energy $E$ to $E_0=0.025$ MeV/u.} 
Consequently, the lower limit for the integration in Eq.\ 
(\ref{eq:stopping_ratio_fluence}) is given by $E_{\mathrm{min}}=0.025$ MeV/u 
having an influence on the STPR of less than 0.00015\% \cite{henkner09}.
A recent review article \cite{belkic10} discusses in some detail the impact 
of electrons in fast ion-atom collisions with respect to hadron therapy as 
well as the possibility to extend \shh\ in a way that also electron 
tracks are considered.
This would allow for studies of the microscopic energy distribution 
in the target medium. 
Tracking of delta-electrons has for example been performed with Geant4 
\cite{taddei08} and a study comparing to the present work might be of 
interest.

\subsection{Analytic expression for the average ion energy}
\label{sec:average_energy}

The stopping-power formula as presented in Eq.\ (\ref{eq:bethe_formula_Fano}) 
for a specified combination of projectile and target is primarily a function 
of the projectile's
kinetic energy which decreases
during the passage through the target medium due to the energy
loss. In order to determine the average energy of the projectiles as a
function of depth a full simulation of the particle transport has to be
performed. This comprises the slowing down caused by all relevant
energy-loss mechanisms including elastic as well as nonelastic
interactions \cite{icru63}. Thereby, nonelastic nuclear reactions produce 
a spectrum of
particles with each particle having an individual energy distribution 
which is furthermore a function of the position in the medium.
%
Consequently, it would be highly desirable to have a simple,
though approximate, analytical expression $E(d)$ for the average energy of the
primary particles with initial energy $E_0$ as function of penetration 
depth $d$.  

Starting with the Bethe formula, but assuming first that the expression in 
the square brackets of Eq.\ (\ref{eq:bethe_formula_Fano}) is independent of 
energy,
$E(d)$ can easily be expressed analytically,
\begin{equation}
 \label{eq:average_energy_simple}
  E(d;E_0,R_{\mathrm{p}}) \approx E_0 \left(
				 1 - \frac{d}{R_{\mathrm{p}}}
				 \right)^{1/2}\,,
\end{equation}
where \rp\ is the practical range. 
\rp\ depends in general on the ion species, $E_0$, and the target material.
For energies relevant in particle therapy \rp\ can often be
approximated by $R_0$ obtained with the continuous slowing down approximation 
(CSDA).

In order to account for the correct energy dependence of the Bethe formula as 
well as nonelastic collisions one has to allow for a more general power-law 
relation,
%
\begin{equation}
 \label{eq:average_energy}
  E(d;E_0,R_{\mathrm{p}}) = E_0 \left(
			     1 - \frac{d}{R_{\mathrm{p}}}
                             \right)^{1/k}\,,
\end{equation}
%
with an exponent $k$. Different values for $k$ are suggested in the literature 
while Kempe and Brahme \cite{kempe08} proposed the use of a dimensionless 
transport parameter $k=E_0 / R_0 S_0$ with $S_0=S(E_0)$. A simple value of 
about $k=1.7$ fits the calculations performed with \shh\ being also 
compatible with \cite{kempe08} and is therefore used in this study.

\subsection{Analytic expression for STPR}
\label{sec:STPR_analytic_expression}

In order to derive an analytic, though approximate, expression of the STPR 
as a function of the depth $d$ for two media $a$ and $b$,
the approximation to the average energy in Eq.\ 
(\ref{eq:average_energy}) can be used together with the ratio of 
stopping powers given in Eq.\ (\ref{eq:stopping_ratio_approximation}). 
Utilizing the non-relativistic relation $v^2 = 2E/m_p$ between the particle 
velocity $v$ and and its kinetic energy $E$, where $m_p$ is the proton mass, 
one obtains the expression
%
\begin{equation}
  \label{eq:STPR_approximation_energy}
  \tilde{S}_{(\mathrm{a}/\mathrm{b})} (d)
  = \frac{\left\langle \frac{Z}{A} \right\rangle _a}
         {\left\langle \frac{Z}{A} \right\rangle _b} \,
    \frac{\ln[E_0/I_{\mathrm{a}}] + C(d)}
         {\ln[E_0/I_{\mathrm{b}}] + C(d)}\,
\end{equation}
where 
\begin{eqnarray}
  \label{eq:STPR_approximation_energy_definition_C}
  C(d)  
        &=& \frac{1}{k}\ln\left[1-\frac{d}{R_p}\right] - 6.1291\, 
\end{eqnarray}
%
%
and $\ln[4\, m_e /m_p] = -  6.1291$ have been used.
%
Similar as in Eq.\ (\ref{eq:stopping_ratio_fluence}) the numerator and
denominator in Eq.\ (\ref{eq:STPR_approximation_energy}) equal except for 
the different $I$-values and $\left\langle Z/A \right\rangle$ ratios. 
%
%
It should be mentioned that in order to keep the expression for 
$\tilde{S}_{(\mathrm{a}/\mathrm{b})}$ as simple as possible its  
derivation has been performed without relativistic kinematics which are 
in principal of relevance for the highest energies used in particle therapy.
Finally, the expression in Eq.\ (\ref{eq:STPR_approximation_energy}) should 
explicitly be formulated for the water-to-air STPR
\begin{equation}
  \label{eq:STPR_approximation_energy_w_a}
  \tilde{S}_{(\mathrm{water}/\mathrm{air})} (d)
  = 1.11195\, 
    \frac{\ln[E_0/I_{\mathrm{water}}] + 1/k \ln\left[ 1 - d/R_{\mathrm{p}}\right] 
                                  + 7.6863}
         {\ln[E_0/I_{\mathrm{air}}] \;\;\,\, + 1/k \ln\left[ 1 - d/R_{\mathrm{p}}\right] 
                                  + 7.6863}\,,
\end{equation}
being the most relevant case for dosimetry in particle therapy, with 
$E_0$ and $I$ in units of MeV/u and eV, respectively. 
For convenience, constants are expressed in numbers, i.e., 
$\ln[4\times 10^6\, m_e /m_p] = 7.6863$ and 
${\left\langle \frac{Z}{A} \right\rangle _{\mathrm{water}}}\,/\,
 {\left\langle \frac{Z}{A} \right\rangle _{\mathrm{air}}} =
0.555076\,/\,0.499189 =
1.11195$ \cite{icru37}.


\section{Results}

In what follows, results for STPRs of pristine Bragg peaks and SOBPs 
are presented. Furthermore, analytic expressions are compared to the 
numerical results obtained with \shh .
For simplicity, only results for ion beams in water are considered, and
STPRs are presented exclusively for water-to-air. 
Furthermore, only stopping-power set 1 of Table \ref{tab:STP_sets} is used 
in order to ensure consistent results except for Fig.\ 
\ref{fig:STPR_C_I}(b) which demonstrates the dependence of 
\swa\ on the choice of the stopping-power set.

\subsection{Pristine Bragg peaks}
\label{sec:results_PBP}


%
\begin{figure}[tb]
 \centering 
 \includegraphics[width=0.48000\textwidth]{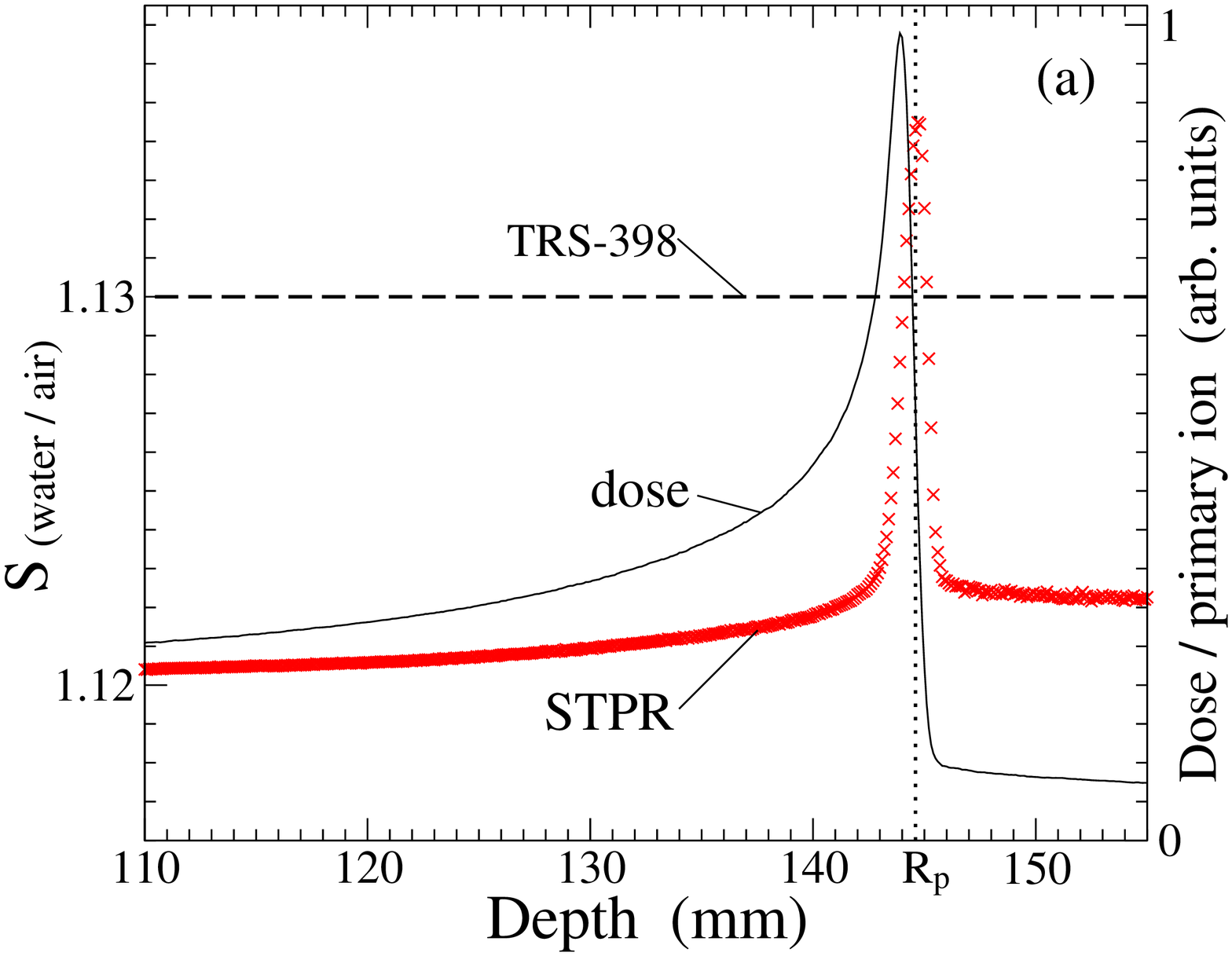}
 \includegraphics[width=0.50000\textwidth]{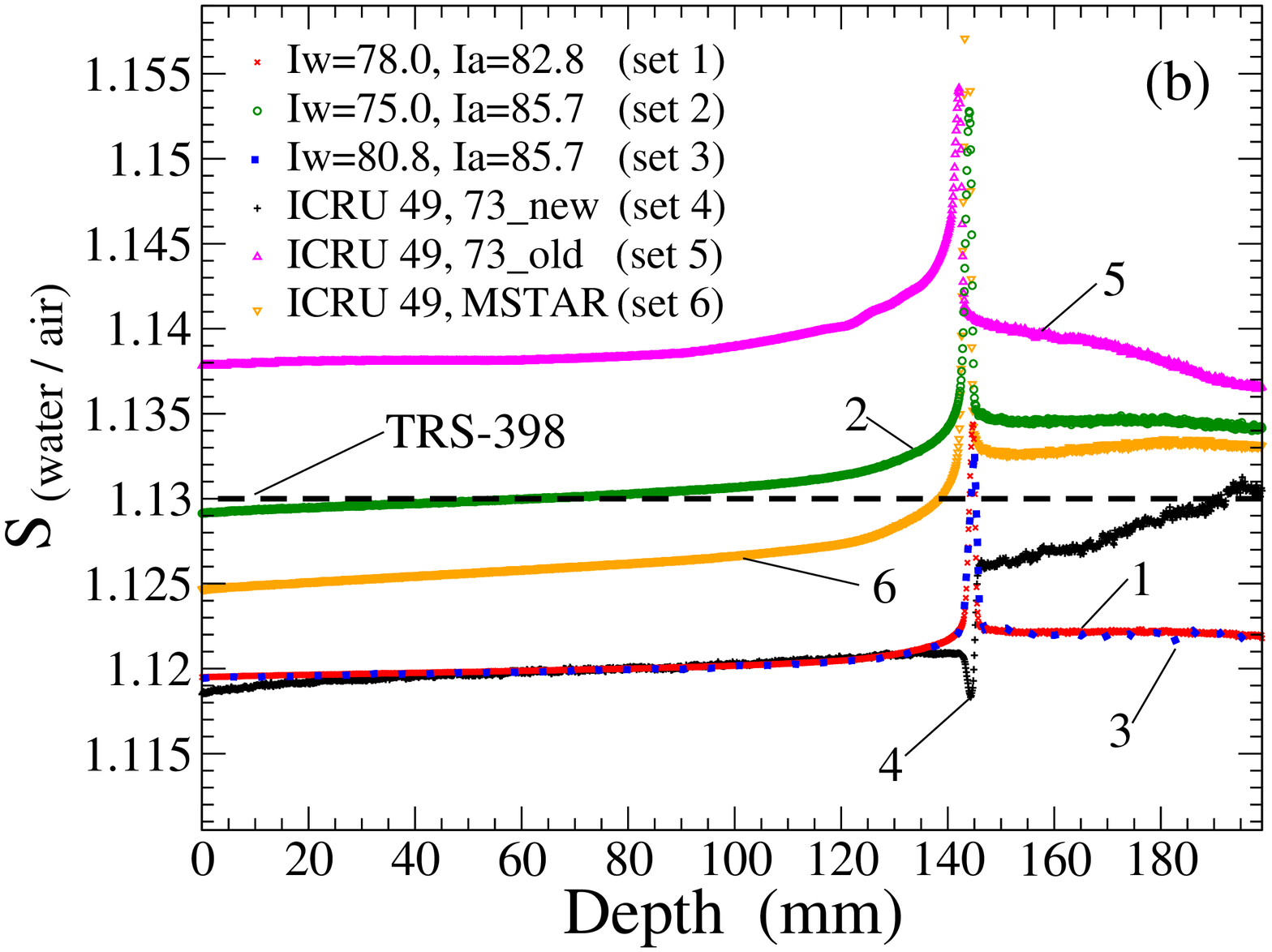}
 \caption{
   270 MeV/u carbon ion beam in water. 
   (a) Depth-dose distribution (line) and stopping-power ratio \swa\ (symbol)
   as a function of depth around the Bragg peak. 
   The stopping-power set 1 is used.
   (b) \swa\ as function of depth obtained with the six stopping-power 
   sets 1\,--\,6 as specified in Table \ref{tab:STP_sets} and 
   the value 1.13 recommended by IAEA in TRS-398 \cite{trs398}.
 }  
  \label{fig:STPR_C_I}
\end{figure}

A comparison between the calculated STPR for water-to-air of 270 MeV/u carbon
ions as a function of depth in
water and the corresponding depth-dose distribution is shown in
Fig.\ \ref{fig:STPR_C_I}(a) focusing on the vicinity of the Bragg peak. 
The maximum of the STPR almost coincides 
with the practical range $R_{\mathrm{p}}=144.6$ mm and therefore appears
to be at a larger depth than the dose maximum. The width of the
STPR peak is considerably smaller than that of the dose curve.
The determined height of the STPR depends to some extend on the finite spatial
resolution along the beam axis. A coarse resolution leads to spatial averaging
and accordingly to a lower peak height of the STPR.
Note, the position of the Bragg peak and therefore \rp\ is only 
influenced by the choice of stopping-power data for water.

The influence on the STPR due to the use of different stopping-power data 
is demonstrated in Fig.\ \ref{fig:STPR_C_I}(b) for a 270 MeV/u carbon pencil 
beam by using all stopping-power sets 1 to 6 of Table \ref{tab:STP_sets}.
In the plateau region, the deviations of the sets 1 to 6 
from the value 1.13 recommended by the IAEA in the
TRS-398 \cite{trs398} is within 1\%.
Set 2, which employs $I_{49}$ for water and air, differs the least. 
This is consistent since the recommendation in TRS-398 is based on the
stopping-power data provided by ICRU report 49.
In contrast to the recommended value 1.13, none of the calculated
STPR curves is constant. However, the relative increase in the plateau 
region up to a depth of 130 mm (\rr\ $\approx 15$mm) is moderate 
and of the order of approximately 0.2\% to 0.3\%.
For all sets of stopping powers, except for set 4, an increase of the STPR
can be observed in the vicinity of the Bragg peak. Set 4 on the other
hand shows a dip. This dip originates mostly from the carbon ions 
and can therefore be attributed to the tabulated data provided by 
in the revision of ICRU 73.

The two STPR curves calculated with set 1 and set 3 --- the latter was used in 
\cite{henkner09} --- lie virtually on top 
of each other in Fig.\ \ref{fig:STPR_C_I}(b) although the $I$-values 
of set 1 and set 3 differ notably.
This can be explained by the differences of the $I$-values for water 
and air which are very similar with 4.8 eV and 4.9 eV for set 1 and set 3, 
respectively. As expected, the \swa\ obtained with the tabulated data from
ICRU 73, that is set 4, agrees with the STPR curve obtained with $I_{73}$, 
set 1, in the plateau region. On the other hand, the STPR curves for the 
sets 1 and 4 deviate around and beyond  \rp .
The use of the stopping-power set 5 by Henkner \ea\ in \cite{henkner09} 
resulted in an unphysical minimum of the STPR in the plateau region.
In the present study, however, no minimum of the \swa\ curve obtained 
with set 5 can be observed.

\begin{figure}[tb]
  \centering
  \includegraphics[width=0.48\textwidth]{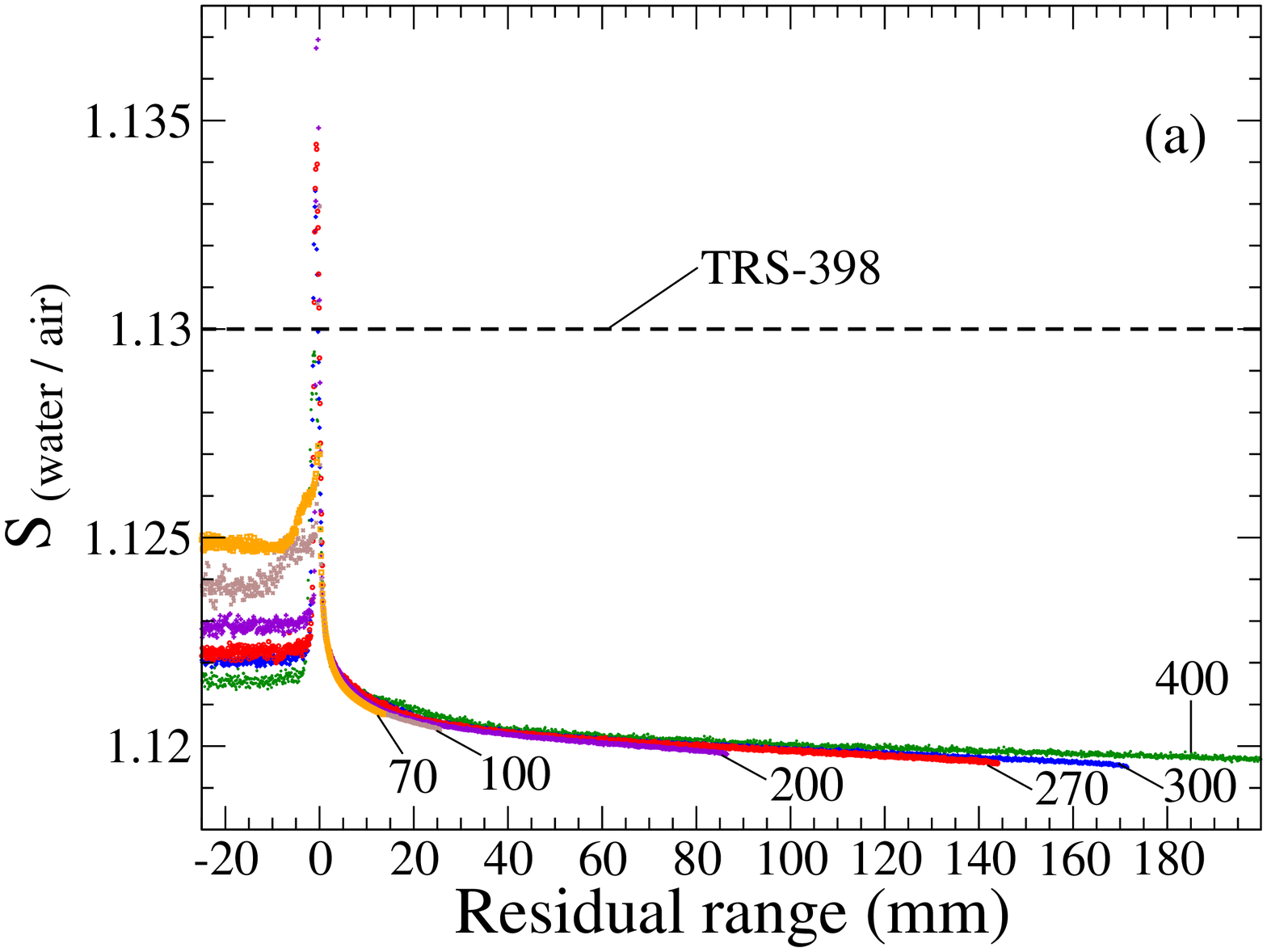}
  \includegraphics[width=0.48\textwidth]{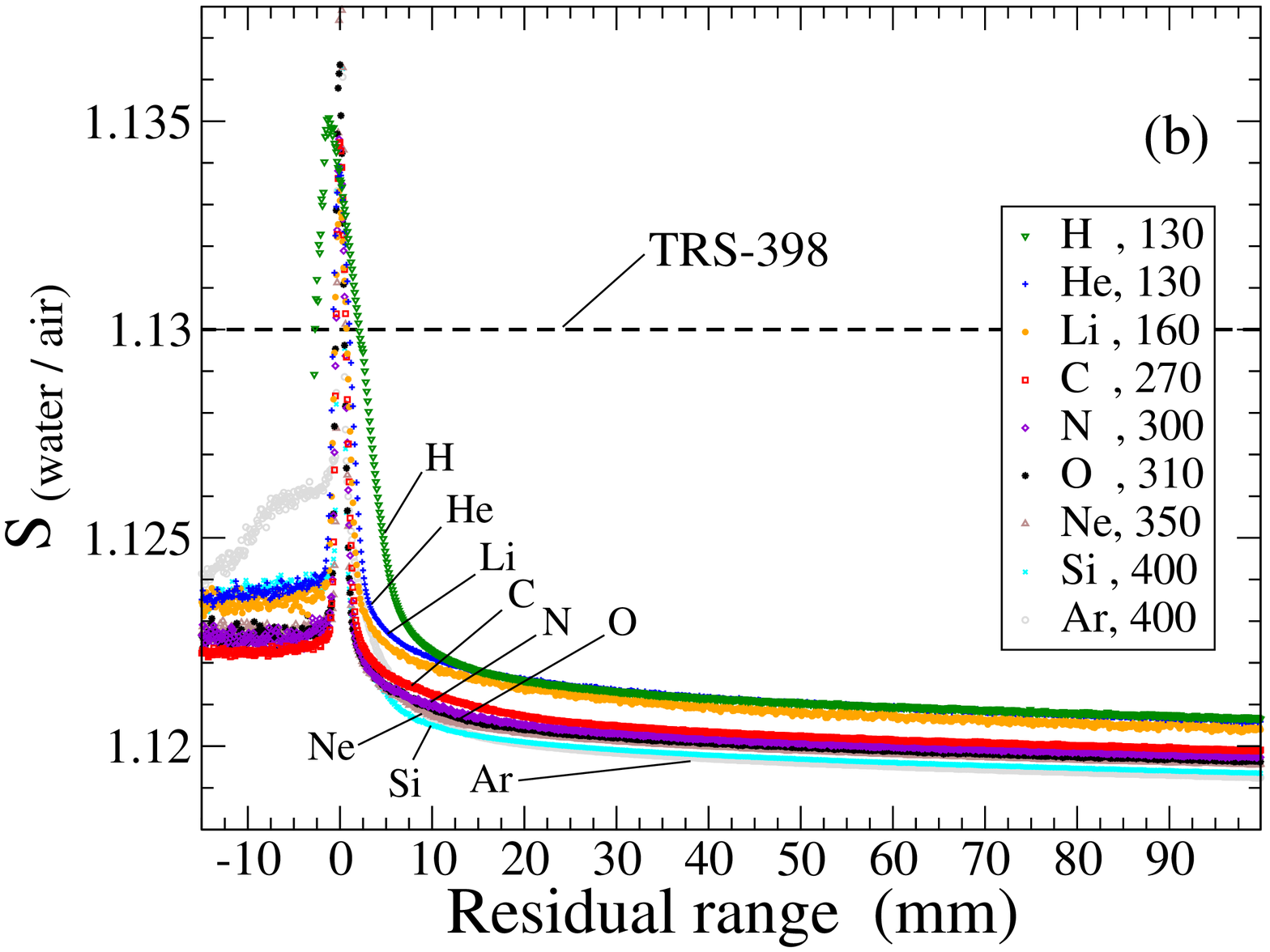}
  \caption{Stopping-power ratio for water-to-air  \swa\ as a function of 
    \rr\ in water. The stopping-power set 1 specified in
    Table \ref{tab:STP_sets} is used. 
    (a) Carbon ions with different initial energies ranging 
    from 70 MeV/u to 400 MeV/u are compared.
    (b) Beams of different ions relevant for particle therapy are compared: 
    H, He, Li, C, N, O, Ne, Si, and Ar. Ion beam energies ranging 
    from 130 MeV/u to 400 MeV/u have been chosen in order to achieve 
    comparable penetration depths.} 
  \label{fig:STPR_C_E} \label{fig:STPR_ION}
\end{figure}

The influence of different initial energies on the STPR for carbon ion beams 
as function of the residual range \rr\ is shown in Fig.\ \ref{fig:STPR_C_E}(a).
The STPR curves as function of \rr\ are almost identical in the plateau region
and therefore independent of the initial energy. 
Around and beyond the Bragg peak the curves are still alike, though 
lower initial energies lead in general to higher STPRs values.


Stopping-power ratio \swa\ as a function of  \rr\ for different ion 
beams relevant for particle therapy: H, He, Li, C, N, O, Ne, Si, and Ar are
presented in Fig.\ \ref{fig:STPR_ION}(b). 
Different beam energies for the individual ions have been chosen, ranging 
from 130 MeV/u to 400 MeV/u, in order to achieve comparable penetration depths.
In the plateau region the STPRs for the different ion beams all share
the same qualitative behavior which has already been discussed before in the 
context of carbon ions, cf.\  Figs.\ \ref{fig:STPR_C_I}(b) and 
\ref{fig:STPR_C_E}(a). 
A comparison among the ions yields that  decreasing STPRs at a given 
\rr\ occur for increasing atomic numbers $z$. 
The decrease becomes less pronounced for larger $z$. 
An exception from this trend is found for H and He ions with very similar
\swa\ in the plateau. 
The relative difference of STPRs between the lightest and heaviest ion 
$z=1$ and $z=18$, respectively, is rather constantly about 0.15\%.

Around and beyond \rp\ (\rr\ $\lesssim 0$), the STPR seems to be larger for 
ions with larger $z$. Although, some dependence also might originate from the 
different initial energies of the ion beams, as has been observed in Fig.\
\ref{fig:STPR_C_E}(a). 
While the STPR for H ions is nearly identical to that for He ions
in the plateau region, differences to all other ions with $z>1$ are
obvious for \rr\ $\lesssim 0$.
However, as discussed in Sec.\ \ref{sec:mm}, the definition
of the practical range \rp\ for protons differs from that used for ions with
$z>1$ which influences --- according to Eq.\
(\ref{eq:residual_range_definition_TRS}) --- also the residual range \rr .

\subsection{Spread-out Bragg peaks}
\label{sec:results_SOBP}

%

%

\begin{figure}[tb]
  \centering
  \includegraphics[width=0.51\textwidth]{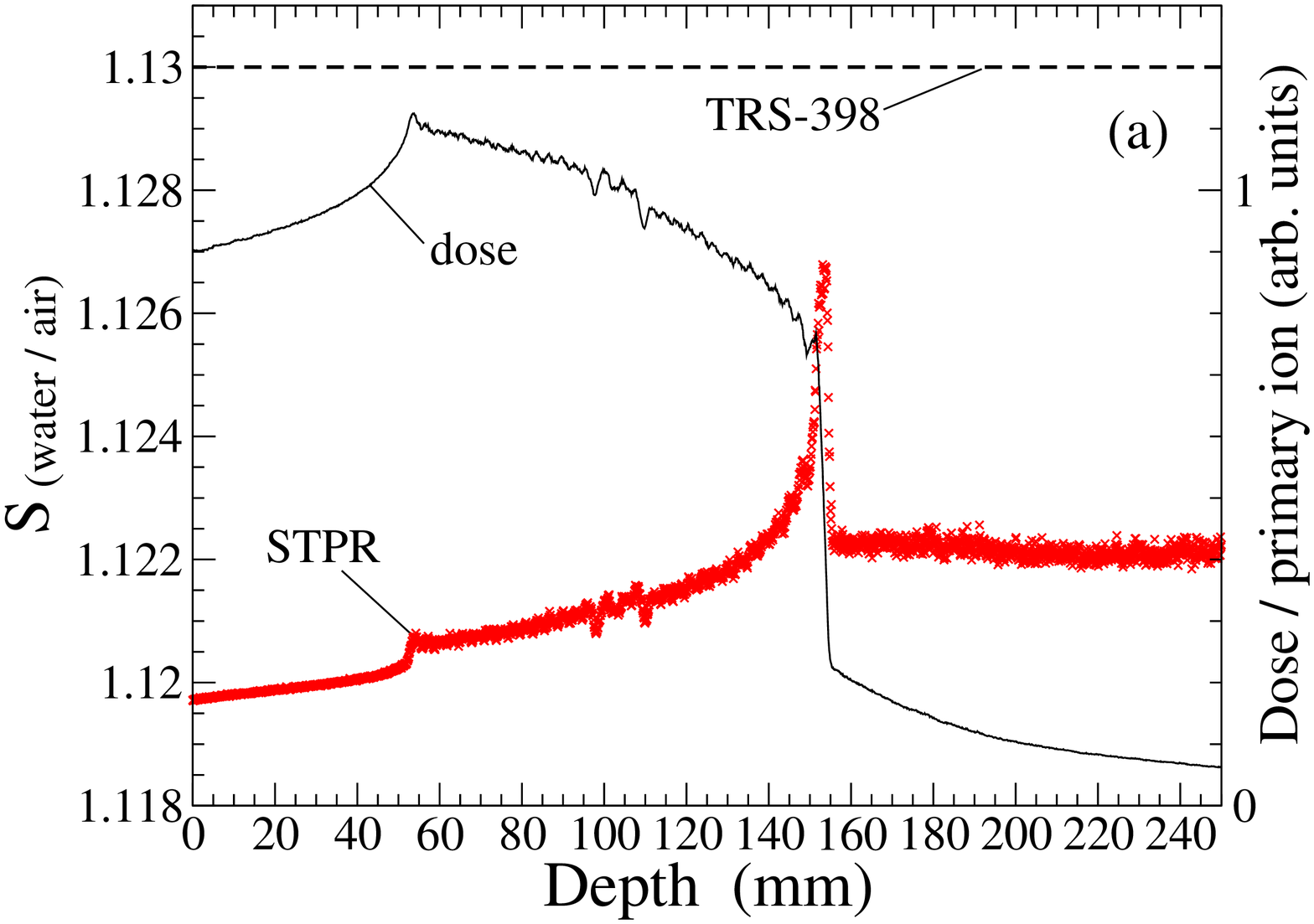}
  \includegraphics[width=0.48\textwidth]{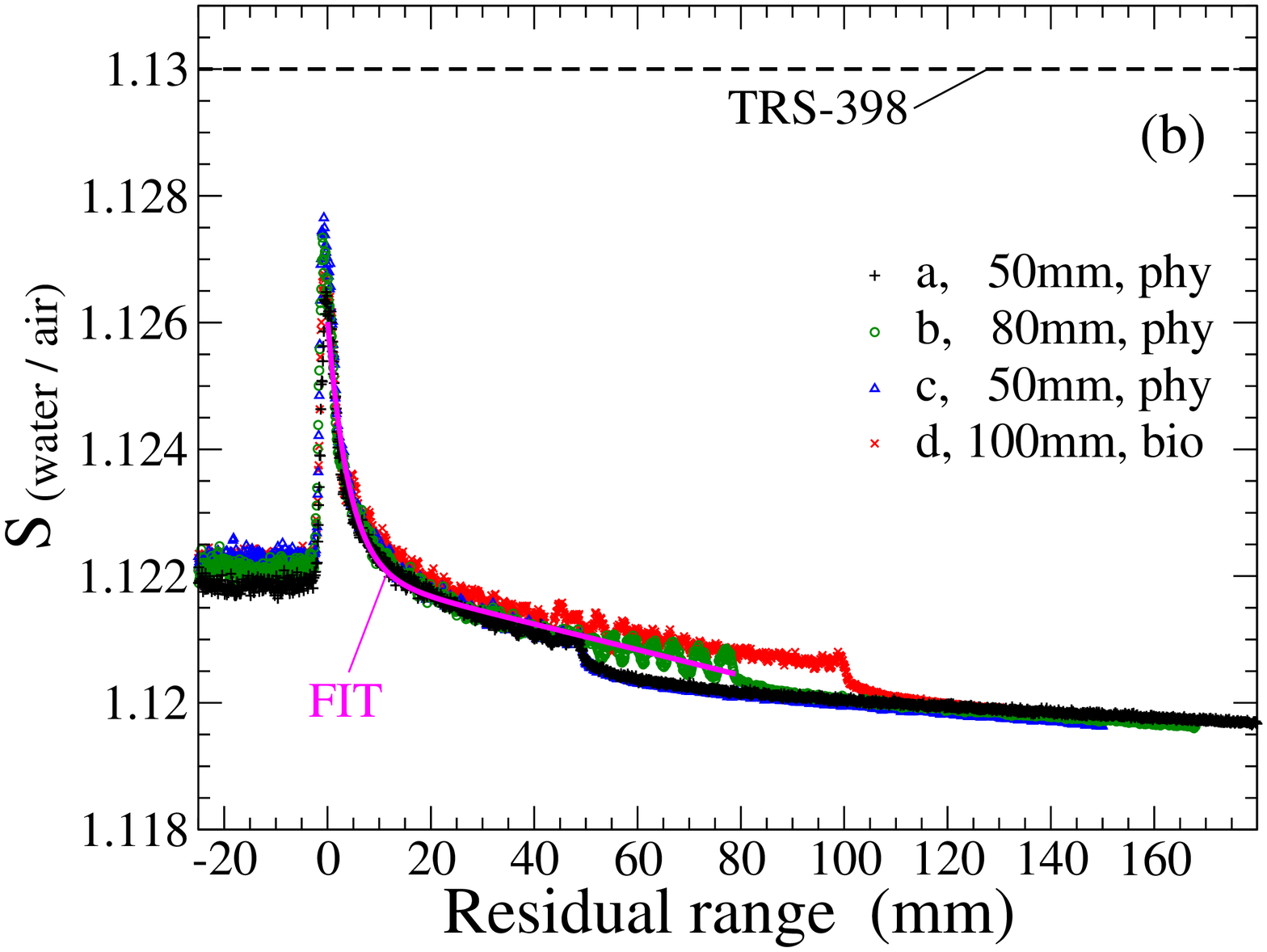}
  \caption{
    Different SOBPs for carbon ions in water as specified in 
    Table \ref{tab:SOBP}. 
    The stopping-power set 1 specified in Table \ref{tab:STP_sets} is used.
    (a) Depth-dose distributions (line) and 
    STPRs \swa\ (symbol) as a function of depth for the biologically optimized 
    SOBP $d$.
    (b) STPRs \swa\ as a function of the residual range \rr\ for all
    four SOBPs $a$, $b$, $c$, and $d$. 
    Also shown is a simple fit for set 1  proposed in 
    Eq.\ (\ref{eq:SOBP_simple_fit_set1}).
  }
  \label{fig:STPR_SOBP}
  \label{fig:STPR_SOBP_residual_range_ICRU}
\end{figure}

Four different SOBPs obtained with carbon ion fields are considered with 
the stopping-power set 1 in this study 
and their properties are listed in Table \ref{tab:SOBP}. 
Figure \ref{fig:STPR_SOBP}(a) displays the depth-dose distribution (line) 
and the corresponding STPR (symbol) of the biologically optimized SOBP $d$.
The proximal start of the SOBP region can be recognized in STPR curve as a 
significant increase. 
Towards the distal end of the SOBP the STPR reveals an exponential increase. 
As for pristine peaks in Fig.\  \ref{fig:STPR_C_I}(a) the maximum of the STPR 
curve is reached close to \rp . 
A sharp fall off of the STPR occurs for depths $d>$ \rp\ which finally 
results in a constant STPR.  
Note, this qualitative description is valid for all SOBPs $a$ to $d$ and is
therefore independent of the specific form of the SOBP or whether it is
physically or biologically optimized.

The STPRs of all four SOBPs $a,\,b,\,c,$ and $d$ are compared in Fig.\
\ref{fig:STPR_SOBP_residual_range_ICRU}(b) as a function of \rr . The
SOBPs $a$ and $c$ share the same width but differ in \rp . 
Therefore, their STPRs as a function of \rr\ are
nearly identical. The STPR curves for $b$ and $d$ show a very similar
behavior as $a$ and $c$ but are extended to 80 mm and 100 mm, respectively,
according to the larger width of their SOBP region.

%

%
It should be mentioned that for SOBP $b$ a number of ripples in the 
STPR can be observed in the proximal SOBP region. They originate from 
the optimization program TRiP which assumes treatment
at the SIS accelerator at Gesellschaft f\"ur Schwerionenforschung (GSI),
Darmstadt, Germany. 
The SIS  provides finite energy steps 
which are more coarse at the lowest energy part when no bolus is applied.

%
\begin{figure}[tb]
  \centering
  \includegraphics[width=0.38\textwidth]{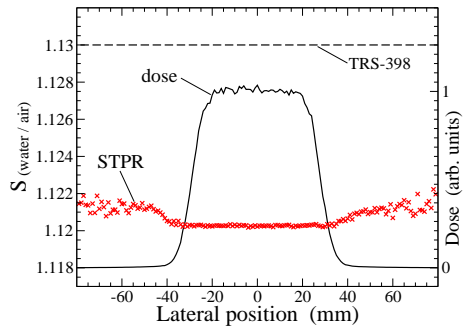}
  \caption{
    SOBP $c$ for carbon ions in water as specified in Table \ref{tab:SOBP}.
    The stopping-power set 1 specified in Table \ref{tab:STP_sets} is used.
    Dose distribution (line) and STPR \swa\ (symbol) as a function of 
    transverse position at a depth $d_{\mathrm{ref}}=150$ mm.} 
  \label{fig:STPR_SOBP_transverse}
\end{figure}

The calculated STPR and dose transverse to the beam axis at the
reference depth $d_{\mathrm{ref}}=150$ mm (defined as middle of SOBP 
\cite{trs398}) is displayed in Fig.\
\ref{fig:STPR_SOBP_transverse} for the SOBP $c$. The STPR is
perfectly constant within the full extension of the SOBP transverse to
the beam axis. A moderate increase of the STPR occurs outside the SOBP.

\subsection{Analytical description of STPR}

The purpose of this section is to relate an analytical description of the STPR
to the numerically obtained STPR. Thereby, it is important to keep in mind 
that according to the results obtained so far
the STPR (a) depends primarily on \rr , that is average ion energy, and (b)
is qualitatively independent of the ion species.

\begin{figure}[tb]
  \centering
  \includegraphics[width=0.48\textwidth]{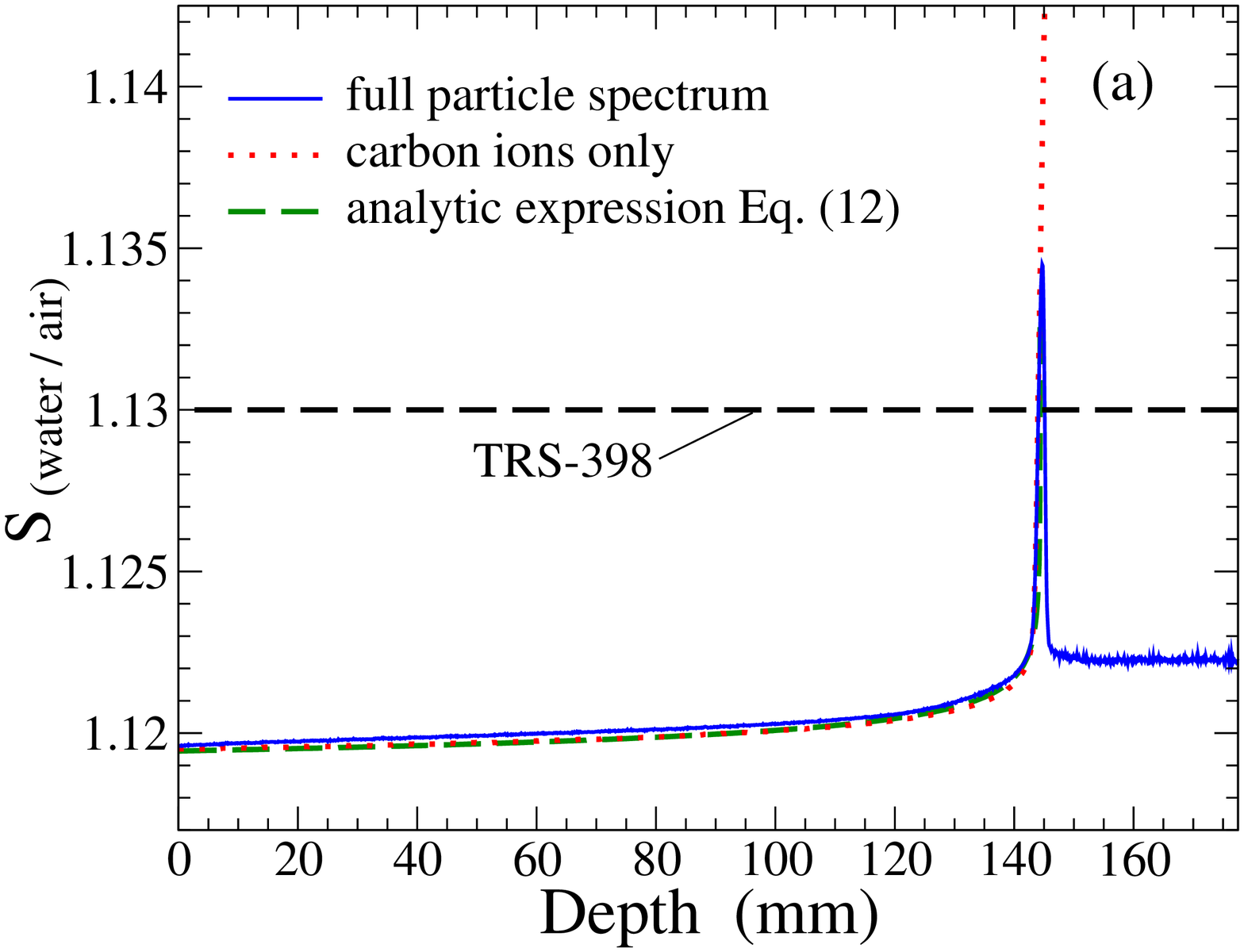}  %
  \includegraphics[width=0.48\textwidth]{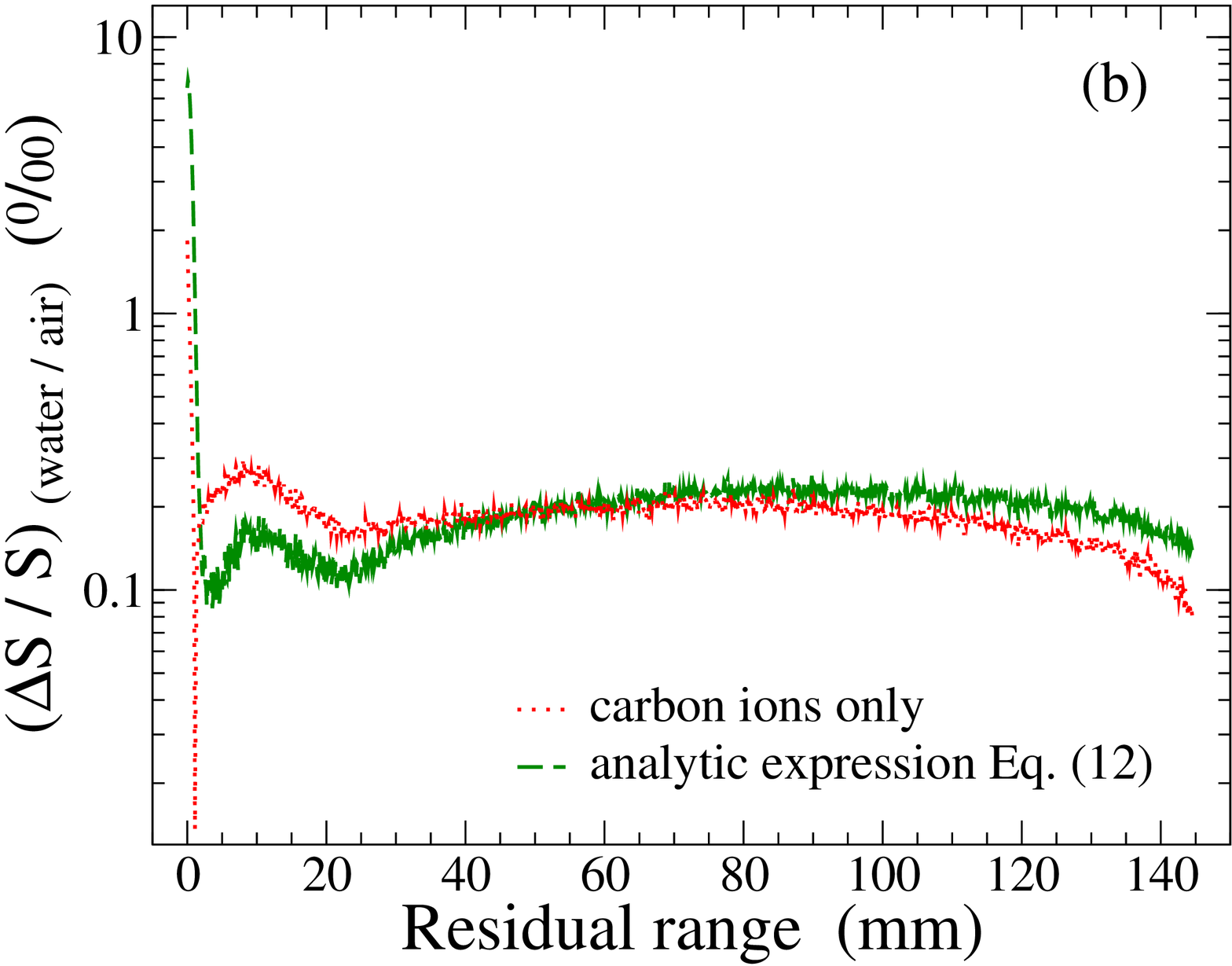}
  \caption{Stopping-power ratio \swa\ for a 270 MeV/u carbon ion beam in water. 
    The stopping-power set 1 specified in Table \ref{tab:STP_sets} is used.
    The STPR obtained with the full particle spectrum is compared to 
    two different approximations:
    STPR determined
    (i) by considering only the contribution from carbon ions, and
    (ii) with the analytic expression proposed in Eq.\
    (\ref{eq:STPR_approximation_energy_w_a}). 
    (a) Absolute \swa\ as a function of depth. 
    (b) Relative difference between the STPR for the
    full particle spectrum and the two approximations to the STPRs 
    presented in (a) as a function of residual range.}    
  \label{fig:STPR_fit}
\end{figure}
%

%

The STPR  \swa\ for a 270 MeV/u carbon ion beam using the stopping-power set 1,
as shown before in Fig.\ \ref{fig:STPR_C_I}(b), which considers the full 
particle spectrum
is compared in Fig.\ \ref{fig:STPR_fit} to two approximations of the STPR: 
%
%
(i) STPR obtained with \shh\ but ignoring the influence of produced fragments 
on \swa\ by only considering the STPR resulting from carbon ions 
and
(ii) the analytic expression in Eq.\
(\ref{eq:STPR_approximation_energy_w_a}) which approximates the STPR with a 
ratio of stopping powers of the primary ions with an average energy depending
on \rr .
%
%
Figure \ref{fig:STPR_fit}(a) clearly shows that especially in the plateau 
region the absolute difference between these three curves is small.
Therefore,  Fig.\ \ref{fig:STPR_fit}(b) additionally shows the relative 
difference $|S^{\mathrm{full}}-S^{\mathrm{appr}}|\,  / \,  S^{\mathrm{full}} $
between the STPR for the \emph{full} particle spectrum, determined according 
to Eq.\ (\ref{eq:stopping_ratio_fluence}), and
the two approximations of the STPR. 
Both approximations reproduce $S^{\mathrm{full}}$ 
within 0.02\% in the whole plateau region. Around \rp\ the difference can be
as large as 1\% and they cannot be applied beyond \rp\ since 
both approximations are based on the primary particles.  
%

%
In principle, the analytical expression for pristine peaks can also be of use
for SOBP. However, since the exact weights of the superposed peaks are not
always known a simple fit as a function of $d$ may be proposed 
\begin{equation}
 \label{eq:SOBP_simple_fit_set1}
  S_{\mathrm{water/air}} (d) = \alpha 
                        + \beta \exp[\gamma (R_{\mathrm{p}} - d)]  
                        + \delta\, (R_{\mathrm{p}} - d)
\end{equation}
which approximates the STPR within the SOBP
region and is also shown in Fig.\
\ref{fig:STPR_SOBP_residual_range_ICRU}(b). The 
values for the four parameters in Eq.\ (\ref{eq:SOBP_simple_fit_set1}) are 

%
  $\alpha=1.12205$, \quad  $\beta=4.0044$E-03, \quad 
  $\gamma=-0.241$, \ and \   $\delta=-2.0238$E-05\,.
%

\noindent These parameters depend on the stopping-power set and slightly on
the ion species \cite{luehr11a}. 
Outside the SOBP region the STPR might be approximated in a
similar way as it is done for pristine peaks considering the average energy
of the primary ions at $d$ which is, however, different from the case of 
pristine peaks. 
%
From  Fig.\ \ref{fig:STPR_SOBP_residual_range_ICRU}(b) it can be observed that 
the fit function proposed in Eq.\ (\ref{eq:SOBP_simple_fit_set1}) clearly is in
acceptable agreement with the \swa\ curves of the four different SOBPs.

\section{Discussion}
\label{sec:discussion}

\subsection{Influence of the inconsistency of ICRU stopping-power 
  data on the STPR} 

It is obvious that the STPR strongly depends on the
stopping-power data which are used as an input for its
determination. 
This applies for \rp\ and accordingly the position of the STPR maximum.
Therefore, the present work focuses mainly on
stopping-power data for water and air which are recommended by ICRU
reports. However, even if one tries to follow these recommendations certain 
inconsistencies still remain and different sets of stopping-power data ---
as listed in Table \ref{tab:STP_sets} --- can be deduced. 
In the case that the tabulated data provided by ICRU are used directly, set 4,
physical quantities of the target media,
e.g., the $I$-value are depended on whether the medium interacts with 
protons and helium ions or with heavier ions. 
Although not explicitly recommended, another approach, which is more 
consistent from a physical point of view, is to use the
$I$-values of only one of the ICRU reports together
with an appropriate stopping formula for all ions, i.e., with $z\le 2$ 
as well as $z>2$.
This is done here for set 1 and set 2 using  $I_{73}$ and $I_{49}$, 
respectively.%
\footnote{A further option is to correct for the $I$-value in one of
the two recommended sets of ICRU tables leading to a more consistent
description of the target media. According to Eqs.\
(\ref{eq:bethe_formula_Fano}) 
this could be done in first 
order using the term  
$0.307075 \,(z^2Z)/(\beta^2 A) \ln[I_{49}/I_{73}]$ for correcting ICRU
report 49 or its negative value for correcting ICRU report 73
\cite{sigmund10a}. This approach is of course not applicable in the
low-energy regime where, on the other hand, the ICRU tables anyhow
provide a limited accuracy only. However, this option has not been
pursued in this work.}  

First, the STPRs in the plateau region ($d<$ \rp ) are discussed. 
Figure \ref{fig:STPR_C_I}(b) reveals that the STPR \swa\ using $I_{49}$
for all ions, set 2, agrees best with a constant
value of 1.13 as recommended in TRS-398 which is plausible since
TRS-398 is based on the data of ICRU 49.
%
\swa\ obtained with $I_{73}$, set 1, is about 1\% smaller compared to that 
obtained with $I_{49}$, set 2
but agrees nicely with the tabulated data recommended by ICRU, set 4.
Second, around \rp\ the STPR curves of all data sets show a distinct 
maximum except for that of set 4 for which a minimum can be observed.
This minimum does not originate 
from differences in the target descriptions due to the use of ICRU 49 and 73 
in set 4 but is exclusively caused by the low-energy ratio of stopping powers 
taken from ICRU 73. 
Third,  beyond the peak ($d>$ \rp )  \swa\ is rather constant for all
stopping-power sets with a consistent description of the target media. 
For set 4 and set 5 the faster decline of the heavier ions relative to the
lighter ones beyond \rp\ leading to a transition from ICRU 73 to 49 is 
clearly revealed by \ref{fig:STPR_C_I}(b).
Therein, the latter two curves finally converge to \swa\ obtained with 
$I_{49}$, set 2.


The use of the out-dated standard, that is stopping tables from ICRU 49
and 73 \emph{without} the revisions of ICRU 73 for water, set 5, is 
not advised. Set 5 yields a STPR which is in
the plateau region about 1\% and 2\% larger than the value 1.13
recommended by TRS-398 and the values obtained \emph{with} revisions of 
ICRU 73 for water, set 4, respectively. 
It should be mentioned, that the results for set 5 in Fig.\ 
\ref{fig:STPR_C_I}(b) do not 
show any unphysical minimum in the plateau region as was observed before in
\cite{henkner09,paul07}. There, the unphysical structure was
attributed to the use of different stopping-power data from ICRU 49
and 73 for different ions. However, in the present work it was
possible to reproduce the exact shape of the curve as shown in  
Fig.\ 2 of \cite{henkner09} by using only three digits of the 
stopping-power data of ICRU 49 and 73
instead of the four digits provided by the tables and therefore to 
contradict the earlier statement. We can conclude that the 
reason for the unexpected behavior observed in the corresponding Fig.\ 2 of
\cite{henkner09} is simply due to errors resulting from a too coarse 
rounding in the applied stopping-power list and are not caused by a 
combined use of ICRU 49 as well as 73.

\subsection{Dependence of the STPR on the average ion energy}

One quintessence of this work is that the STPR for a given set of
stopping-power data is mostly determined by the average energy of the
ions, rather than their initial energy or their charge 
which is nicely confirmed by Fig.\ \ref{fig:STPR_C_E} 
Figure \ref{fig:STPR_fit} shows furthermore that for $d<$ \rp\ 
the STPR
is completely dominated by the STPR of the primary ion species and the
relative deviation is of the order of 0.02\%.
In a next step, the STPR of the primary ions can be nearly exactly reproduced
using the ratio of stopping powers, as expressed in Eq.\ 
(\ref{eq:stopping_ratio_approximation}), for the average energy of the primary
ions at a depth $d$. Since the average energy can be rather accurately 
expressed as a function of $d$, in a final step this energy function is used
together with Eq.\ (\ref{eq:stopping_ratio_approximation}) to 
formulate the analytical expression of the STPR in 
Eq.\ (\ref{eq:STPR_approximation_energy_w_a}). The analytical expression
reproduces the STPR of the primary ions very well and deviates accordingly
also in the order of 0.02\% from the correct STPR obtained with the complete 
particle spectrum.


Note, the key advantage of the proposed analytical expression is its
flexibility. 
%
It is not restricted to a specific set of stopping-power data. 
Consequently, it could be easily adopted to any new recommendation by ICRU. 
%
It is not restricted to specific primary ions and only  
their average energy as a function of depth is required.
%
It is not 
restricted to a specific combination of target materials such as water and 
air being in the focus of the present work. For example, it can be 
straightaway used for STPR for water to tissue and air to tissue which 
would be of interest when comparing dose to medium with dose to water as 
recently discussed by Paganetti \cite{paganetti09}.


These findings lead to two central insights. 
First, in contrast to the
presumption in TRS-398, the knowledge of the whole particle spectra 
is not of practical importance for the plateau and peak peak region.%
\footnote{The relevance of the whole particle spectra for determining
the STPR increases if for some secondary particles different
stopping-power data are employed, e.g. ICRU 73 and 49.} 
The STPR can be simply approximated with a
relative deviation much smaller than the uncertainties of all  
available stopping-power data. On the other hand, the secondary
particles are of central importance beyond 
\rp\ where the primary ions are ceased. 
Second, it is not necessary to study STPRs for all ions
independently since they can be approximated in the same way
as proposed in Eq.\ (\ref{eq:STPR_approximation_energy}). 
The small quantitative differences among the STPRs of the
various ions shown in Fig.\ \ref{fig:STPR_ION}(b) may be explained best
with the different average energies at a given \rr .

%
%

%
\subsection{STPRs for SOBPs}

Clinical applications require a relatively uniform dose to be delivered
to the volume to be treated and for this purpose the ion beam has to be
spread out both laterally and in depth.
With respect to the STPR of SOBPs the gist of this work is that the
qualitative behavior of the STPR is hardly dependent on the specific
spatial form of the SOBP, its depth in the medium, and whether it is
optimized for a homogeneous physical dose or relative biological dose.

This statement is nicely verified in Fig.\
\ref{fig:STPR_SOBP_residual_range_ICRU}(b) where \swa\ is displayed
as a function of \rr\ for four SOBPs specified in
Table \ref{tab:SOBP}. In order to understand the observed uniform
behavior of the STPR one has to keep in mind that a SOBP is a
superposition of Bragg peaks of different intensities and \rp\ which
is usually constructed from the distal end, i.e.\ $R_{\mathrm{res}}=0$,
toward the proximal start. Consequently, the properties of the SOBP up
till a residual range \rr\ are only weakly influenced by the
properties of the SOBP for larger \rr .
It is therefore possible and practical to propose a fit to
the STPR \swa\ for the SOBP region, as it is done in Eq.\
(\ref{eq:SOBP_simple_fit_set1}) for carbon ions and set 1.
%
%
A general drawback of the fit function compared to an analytical expression
is that the parameters of the former explicitly depend on the ion species
and stopping-power data.
%
In analogy to the findings for pristine peaks quantitative differences can 
be expected for ion species other than carbon ions.
However, a detailed study of a number of other ions is beyond the scope of 
this article and might be addressed elsewhere \cite{luehr11a}.
The qualitative dependence of \swa\ on the stopping-power set is similar 
as discussed before for the pristine peaks.
Accordingly, the quantitative difference between \swa\ obtained with set 1 
and set 4 is 
due to the two features
observed for set 4, namely, the minimum at \rp\ and the strong increase of 
the STPR beyond \rp\ caused by the ICRU 49 tables for protons and helium.

In TRS-398 reference conditions for the determination of absorbed dose
for ion beams are specified. The reference depth $d_{\mathrm{ref}}$ for 
calibration should
be taken at the middle of the SOBP, at the center of the target volume.
It can be seen in Fig.\ \ref{fig:STPR_SOBP_transverse} that a
positioning error of a dosimeter transverse to the beam axis has no
relevance on the STPR as long as the position is within the SOBP. This
is plausible since the average energy of the
ions should be same at the same depth.
A misalignment along the beam axis, on the other hand, may have an
influence as seen in Fig.\ \ref{fig:STPR_SOBP_residual_range_ICRU}. The
influence is largest for a SOBP with small width, for which the gradient of 
STPR is largest, and becomes smaller for large widths.
Therefore, an extended SOBP might be recommended for accurate dose 
measurements in a practical quality-assurance setting. 
The total variation of the STPR along the beam axis observed in Fig.\
\ref{fig:STPR_SOBP_residual_range_ICRU}(b) 
is of the order of 0.8\%.

\section{Conclusions}

Calculations of the water-to-air stopping-power ratio (STPR) \swa\
using the Monte Carlo transport code \shh10A are performed for different ions 
in a range of $1\le z \le 18$.
The STPR is determined on-line considering the track-length fluence spectra 
of all primary and secondary particles as recommended by IAEA in TRS-398. 
In addition to providing accurate quantitative results the focus of this
work is put on a thorough qualitative understanding of the dependencies of
the STPR and the relevance for particle therapy.
%

STPRs obtained with different sets of stopping-power data 
recommended by ICRU \cite{icru49,icru73}, including the very recently 
revised data for water \cite{icru73a}, are compared with the value 1.13 
recommended for \swa\ in 
TRS-398 \cite{trs398} resulting in deviations of the order of 1\% in the 
plateau region. 
The change of the STPR due to the contribution of secondary particles is only
of the order of 0.02\% for pristine peaks in the plateau region and up to 
the Bragg peak.
It can be shown that for a given set of stopping-power data the STPR at a 
residual range \rr\ is mostly determined by the average energy of the primary 
ions, rather than their initial energy or their charge $z$.
A convenient analytical expression for the STPR as a function of depth in 
water is proposed for the plateau region up to the Bragg peak
which deviates in this region by about 0.02\% from the obtained results 
for \swa .
The most valuable property of the analytical formula is its flexibility. 
It is in principle not restricted to any specific ion, 
stopping-power data, combinations of target media, or initial ion energies. 
For the case of spread-out Bragg peaks (SOBPs) it can be concluded that 
the qualitative behavior of the STPR is hardly dependent on the specific
spatial form of the SOBP, its depth in the medium, and whether it 
provides a homogeneous physical dose or relative biological dose.
A fit function is provided to approximate the STPR within the SOBP region 
for carbon ions.

Finally, it can be stated that no further theoretical studies of STPRs 
heading only for higher accuracy are expedient, as long as no consistent 
set of  relevant stopping-power data for all ions is recommended, preferably 
with smaller uncertainties. 

\ack
This work is supported by the Danish Cancer Society (http://www.cancer.dk), and 
the Lundbeck Foundation Centre for Interventional Research in Radiation 
Oncology (http://www.cirro.dk).
\section*{References}

\bibliographystyle{unsrt}

\end{document}